\documentclass[twocolumn]{aa}
\usepackage{appendix,float}
\usepackage[T1]{fontenc}
\usepackage[utf8]{inputenc}
\usepackage{lmodern}
\usepackage{graphicx,natbib,journal_names}
\usepackage[breaklinks,colorlinks,urlcolor=blue,citecolor=blue,linkcolor=blue]{hyperref}
\usepackage{multirow,afterpage,pdflscape,lipsum,capt-of,mathtools}

\usepackage[dvipsnames]{xcolor}
\usepackage{comment}
\usepackage{float}
\usepackage{txfonts}
\usepackage{soul}
\usepackage{array}
\usepackage{hyperref}
\usepackage{amsmath}
\usepackage[switch]{lineno}
\usepackage{xcolor}
\usepackage{multicol}

\begin{document}
\title{Near-Sun Switchbacks Show Statistical Signatures of Solar Activity }
\titlerunning{Near Sun Switchbacks}
\author{Sneha Pandit\inst{1}, Thierry {Dudok de Wit}\inst{2,3}, Clara Froment\inst{3}, Durgesh Tripathi\inst{1}, Vishal Upendran\inst{4,5}, Gabriel Ho Hin Suen\inst{3}}
\institute{Inter-University Centre for Astronomy and Astrophysics, Post Bag 4, Ganeshkhind, Pune 411007, Maharashtra, India
\and
International Space Science Institute, ISSI, Bern, Switzerland
\and
LPC2E, OSUC, Univ Orleans, CNRS, CNES, F-45071 Orleans, France
\and
SETI Institute, Mountain View, CA, USA - 94043
\and
Lockheed Martin Solar and Astrophysics Laboratory, Palo Alto, CA, USA - 94304
\\
\email{sneha.pandit@iucaa.in}}
\authorrunning{S. Pandit, T. Dudok de Wit, C. Froment, D. Tripathi, V. Upendran, G. Suen}
\abstract
  {
The large amplitude, Alfvénic deflections in the solar wind, called magnetic switchbacks, are a ubiquitous feature of the inner heliosphere, yet their origin and variability remain poorly understood.
  } 
    {We investigate the dependence of large amplitude magnetic fluctuations in the inner heliosphere on both heliocentric distance and global solar activity.}
  {
   Using Parker Solar Probe observations, we quantify switchback deflections through their normalised deflection angle $z$. We examine both their distribution and the probability of large events, combining radial binning, sunspot-number-based activity classification, and regression analysis.
}
  {
 We find a statistically significant but weak dependence of switchback properties on solar activity, with a decrease in large deflections at higher activity levels, alongside a modest increase with heliocentric distance. The weak activity trend suggests that multiple processes act simultaneously. Solar-cycle variations in coronal magnetic topology may modulate switchback generation at the source, while nonlinear in situ effects in the solar wind may partially evolve these signatures.}
   {   Our results indicate that switchbacks retain only a limited imprint of solar activity, reflecting a coupled interplay between coronal origin and in situ evolution. This study provides a quantitative framework to disentangle activity and radial effects, and highlights the need for multi-parameter and multi-spacecraft analyses to fully understand the origin and evolution of switchbacks.}
    \keywords{Sun: Heliosphere, Solar Wind, Activity, Methods: Statistical}

\maketitle

\vspace{-20pt}
\section{Introduction}

Switchbacks observed in situ in the solar wind are localised magnetic deflections in which the radial field  changes its direction temporarily. They have emerged as 
{important structuring features} in the young solar wind, with implications for its acceleration, turbulence, and magnetic connectivity. Understanding their origin is therefore central to broader questions in heliophysics, including how the corona structures and releases the solar wind. 

Two main classes of mechanisms have been proposed for the formation of switchbacks: processes rooted in the solar atmosphere, such as interchange reconnection and surface motions~\citep[e.g.][]{upendran_2021, upendran_2022} and in situ processes, such as nonlinear Alfvénic evolution and flux-tube expansion \citep[e.g.][]{Squire2020,Mallet2021}. These scenarios are not mutually exclusive. For recent review on switchbacks, see \citet{2026SSRv..222...14B} and \citet{2026SSRv..222...43W}. Current understanding suggests that switchbacks may be seeded in the corona and subsequently modified by in situ evolution as they propagate outward, shaping the properties of the young solar wind inside $\sim0.3$ AU observed by Parker Solar Probe \citep[PSP,][]{fox_solar_2016}. Despite recent advances, it remains unclear to what extent switchback occurrence and properties are controlled by the large-scale coronal magnetic topology.

Recent numerical simulations provide strong support for a coronal origin, demonstrating that interchange reconnection, jets, and the release of twisted magnetic structures can generate switchback-like perturbations that survive propagation into the heliosphere \citep[e.g.,][]{wyper_imprint_2022, 2024A&A...692A..71T}. However, establishing this link observationally remains challenging. Case studies relying on magnetic connectivity suffer from uncertainties in field line tracing and coronal mapping. In addition, statistical approaches are limited by a gap between the remote-sensing and in situ observations, leading to the difficulty of isolating source region signatures \citep[e.g.,][]{bizien_tracing_2025}.

All these mechanisms, and especially the ones that involve in situ generation, are likely to depend on the solar wind parameters. There is clear evidence that the solar wind speed \citep{2023ApJ...944...82S} and distance from the Sun \citep{Tenerani_2021} play an important role. Here we consider one parameter that has been overlooked so far, namely the level of solar activity, and investigate what this tells about the origin of switchbacks. We consider the sunspot number as a proxy for {the level of} solar activity.

Two obvious characteristics of switchbacks for exploring the impact of the level of solar activity are their occurrence rate (number of events per unit time or per unit length) and their duration. However, both of these properties are highly dependent on the  motion of the spacecraft relative to that of the solar wind. Indeed, when the spacecraft is corotating with the Sun, it samples a population of switchbacks that are propagating in time, whereas otherwise it rather performs a spatial cut. Several studies \citep{Laker2021,laker_switchback_2022,Horbury2020} have shown how much inferred switchback properties can be affected by these. For that reason, we focus here on the deflection angle with respect to the Parker spiral. This angle is most likely determined by the generation mechanism and by the outward propagation of the switchback and, to the first order, is invariant to the relative motion of the spacecraft.

In the following, we investigate how the distribution of deflection angles depends on the distance from the Sun and the level of solar activity. For a Sun-origin of switchback due to interchange reconnection, we are investigating the impact of solar activity on the amount of open-close flux interaction. For in situ origin, we are investigating the impact of solar activity on in situ generation and propagation of switchbacks. Data processing and statistical methodology are described in Sect.~\ref{Sec:Methods}. Sect.~\ref{Sec:Results} presents the quantitative trends, followed by a discussion and conclusions in Sect.~\ref{Sec:Discussion}.

\vspace{-15pt}
\section{Description of the switchback dataset}
\label{Sec:Methods}

\begin{figure}[htp!]
\centering

\vspace{-15pt}
\includegraphics[width=0.5\textwidth, keepaspectratio]{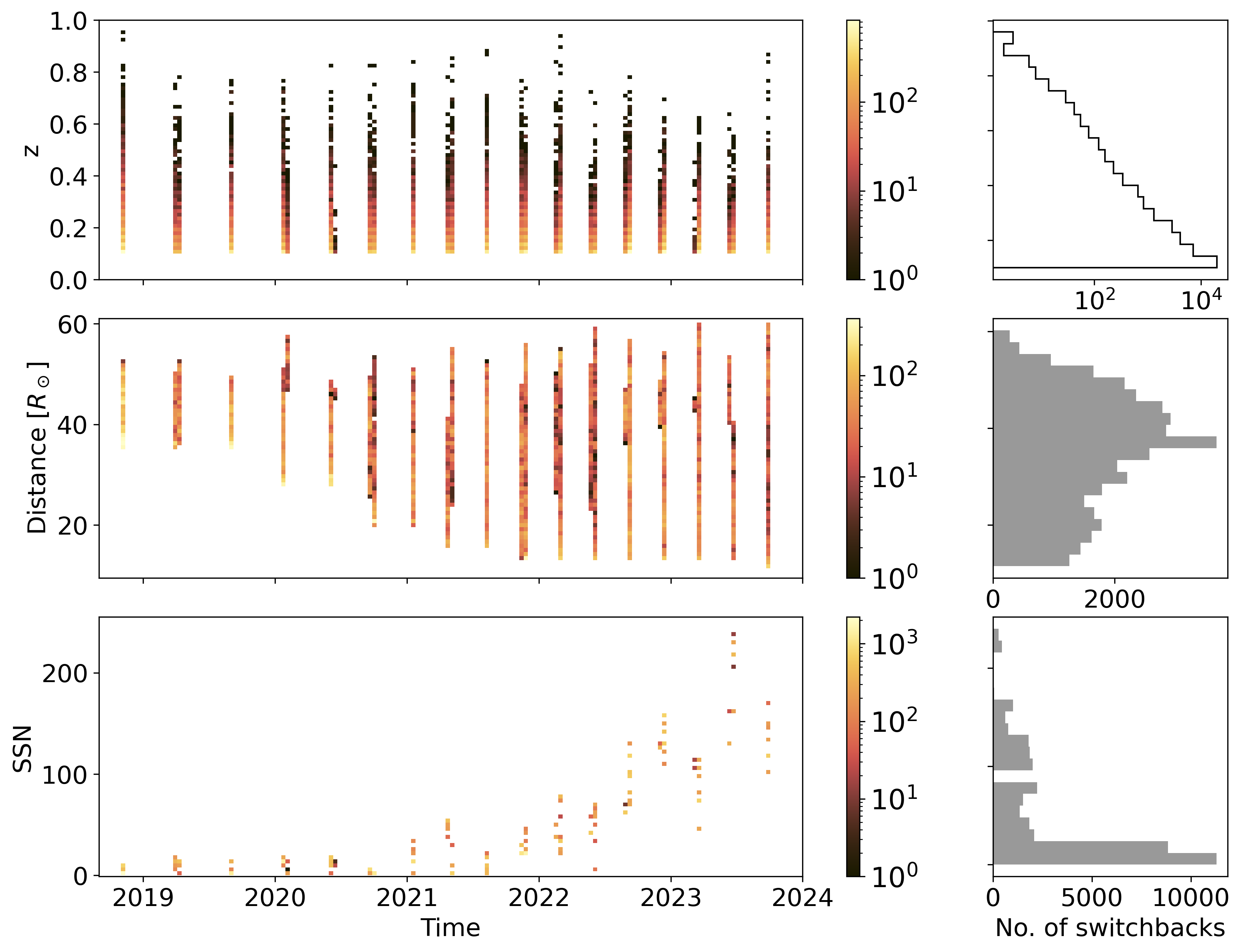}
\caption{Distribution of the switchbacks dataset: top panel shows the  distribution of the z-parameter along with histogram of occurrences of switchbacks with logarithmic scale. The middle panel shows the time series of the distance of PSP from the Sun along with histogram of the distribution of the distances. The bottom panel shows the time evolution of the sunspot number (SSN) along with histogram of SSN distribution of the right panel. All share the time axis for better comparison. Colourbars for the left panels show the density of datapoints.}
\label{fig:description_of_dataset}
\end{figure}

The switchbacks in PSP observations are defined using the vector magnetic field data from the MAG/FIELDS instrument \citep{bale_fields_2016} onboard PSP. We use the $z$ parameter, which is the normalised deflection angle and is defined as 
\begin{equation}
    z = \frac{1}{2} (1-\cos \theta) \ .
\end{equation}
Here, $\theta$ is the angle between the observed magnetic field and the estimated Parker spiral. To detect individual switchbacks, we first determine the orientation of the Parker spiral as in \citep{dudok2020}, following which we compute the deflection of the spiral and its normalised version, the $z$ parameter. Next, we use an algorithm that detects pulse-shaped deflections in $z$ to identify {individual} switchbacks.

\begin{figure*}[ht!]
\centering
\vspace{-15pt}
\includegraphics[width=1.0\linewidth]{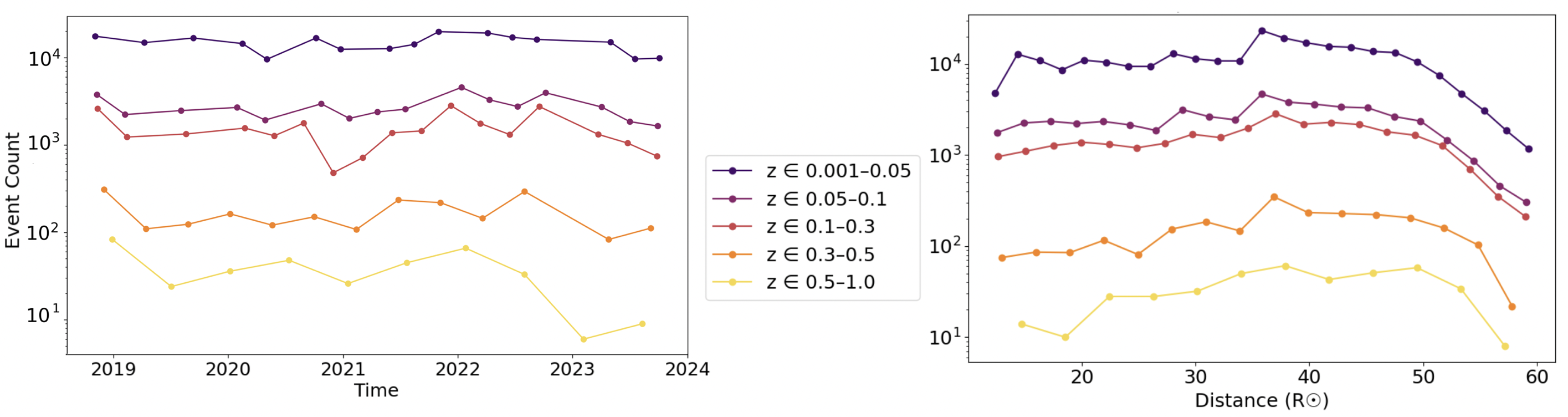}
\caption{Evolution of the number of {switchback} events per unit time (left panel) and distance (right panel) per $z$ bin. The bins are logarithmically spaced (0.0–0.05, 0.05–0.1, 0.1–0.3, 0.3–0.5, 0.5–1.0) as indicated in the legend in the middle.}
\label{fig:z_binned_time_distance}
\end{figure*}

In Fig.\ref{fig:description_of_dataset}, we display our switchback dataset, which covers the first 17 solar encounters of PSP, from 2018 to 2024. {This excludes the times when {Coronal Mass Ejections} (CMEs) reach the PSP locations based on CME speed and PSP trajectory.} In the following, we only consider switchbacks with deflections satisfying $z>0.1$ since smaller ones can be difficult to distinguish from stochastic fluctuations \citep{2026SSRv..222...14B}. In Fig.~\ref{fig:z_binned_time_distance} only, we shall briefly include smaller deflections with $z<0.1$. 

In the left column of Fig.\ref{fig:description_of_dataset} we show the temporal evolution of key quantities of interest for the 17 encounters. Their histograms are displayed in the right column. The top row shows the $z$ parameter, with a distribution that is highly skewed towards small values. The middle row shows the radial distance of PSP from the Sun. Early encounters had perihelia located at 20-35 solar radii (Rs). Subsequent encounters came closer to the Sun, with perihelia as low as 11 Rs. The histogram shows that the distances are not evenly represented, which will be one of the challenges faced by our study.

Finally, the bottom panels show the daily sunspot number {distribution} \citep{2010LRSP....7....1H, clette_recalibration_2023}\footnote{\href{https://www.sidc.be/SILSO/datafiles}{https://www.sidc.be/SILSO/datafiles}} sampled at the times of the switchbacks. The sunspot number is a classical proxy for the level of solar activity. Our dataset starts near solar minimum, showing a gradual increase in solar activity. {Because of that, the majority of observations are made at periods of low solar activity.} Note that the sunspot number here represents an average over one Carrington rotation.

Figure~\ref{fig:z_binned_time_distance} shows the evolution of the $z$ parameter {versus time (left plot) and versus radial distance from the Sun (right plot)}. We divided the data set into 5 bins of equiprobable values of $z$, using Doane’s formula \citep{Doane01111976} to ensure that each bin contains a sample of sufficient size to calculate meaningful statistical quantities.

The temporal evolution (left plot) does not reveal any clear trend, presumably because each data point mixes observations made at different distances. However, a clear non-monotonic evolution appears when the deflections are expressed versus distance, showing in particular that large and small deflections do not behave exactly in the same way. We shall further statistically test these non-linear evolutions that are suggested by Fig.~\ref{fig:z_binned_time_distance}.

\vspace{-15pt}
\section{Results}
\label{Sec:Results}
\subsection{Switchback Activity Metrics}

Figure~\ref{fig:z_binned_time_distance} suggests that while the distribution of deflections varies in a rather complex way with distance and time, the relative proportion of small deflections shows a more systematic pattern. For that reason, we introduce the Small z Ratio ($SzR$),

\vspace{-7pt}
\begin{equation}
    \text{$SzR$}=\frac{n(0.1<z<0.2)}{n(z)}.\label{equation:$SzR$}
\end{equation}

This index expresses the relative proportion of small deflection switchbacks.

\begin{figure}[ht!]
\centering
\includegraphics[width=0.5\textwidth, height = 1.0\textheight, keepaspectratio]{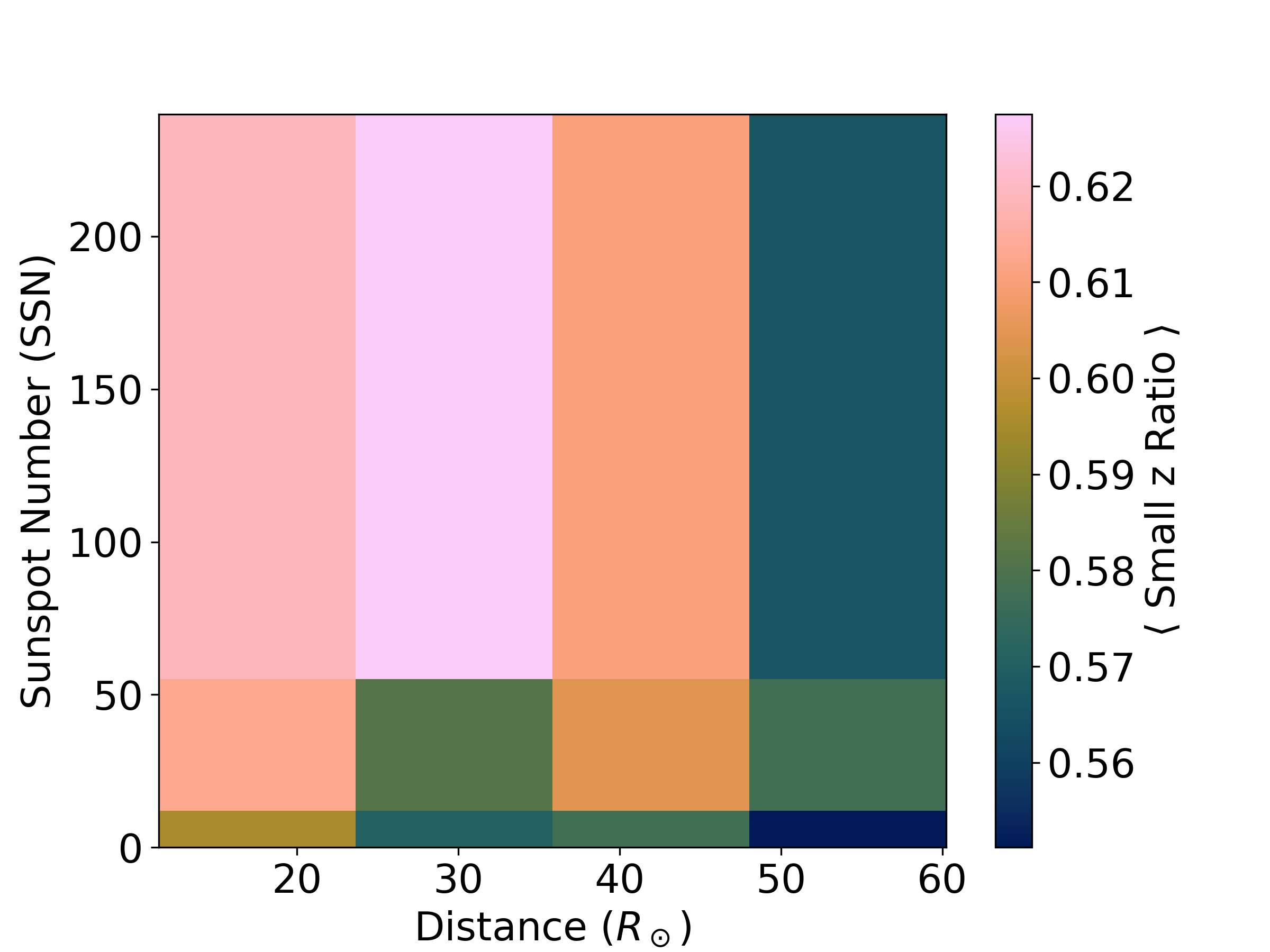}
\caption{The mean Small z Ratio ($SzR$) binned in distance and SSN.}
\label{fig:AS1_phase_distance}
\end{figure}

\begin{figure}[ht!]
\centering
\vspace{-15pt}
\includegraphics[width=0.5\textwidth, height = 0.88\textheight, keepaspectratio]{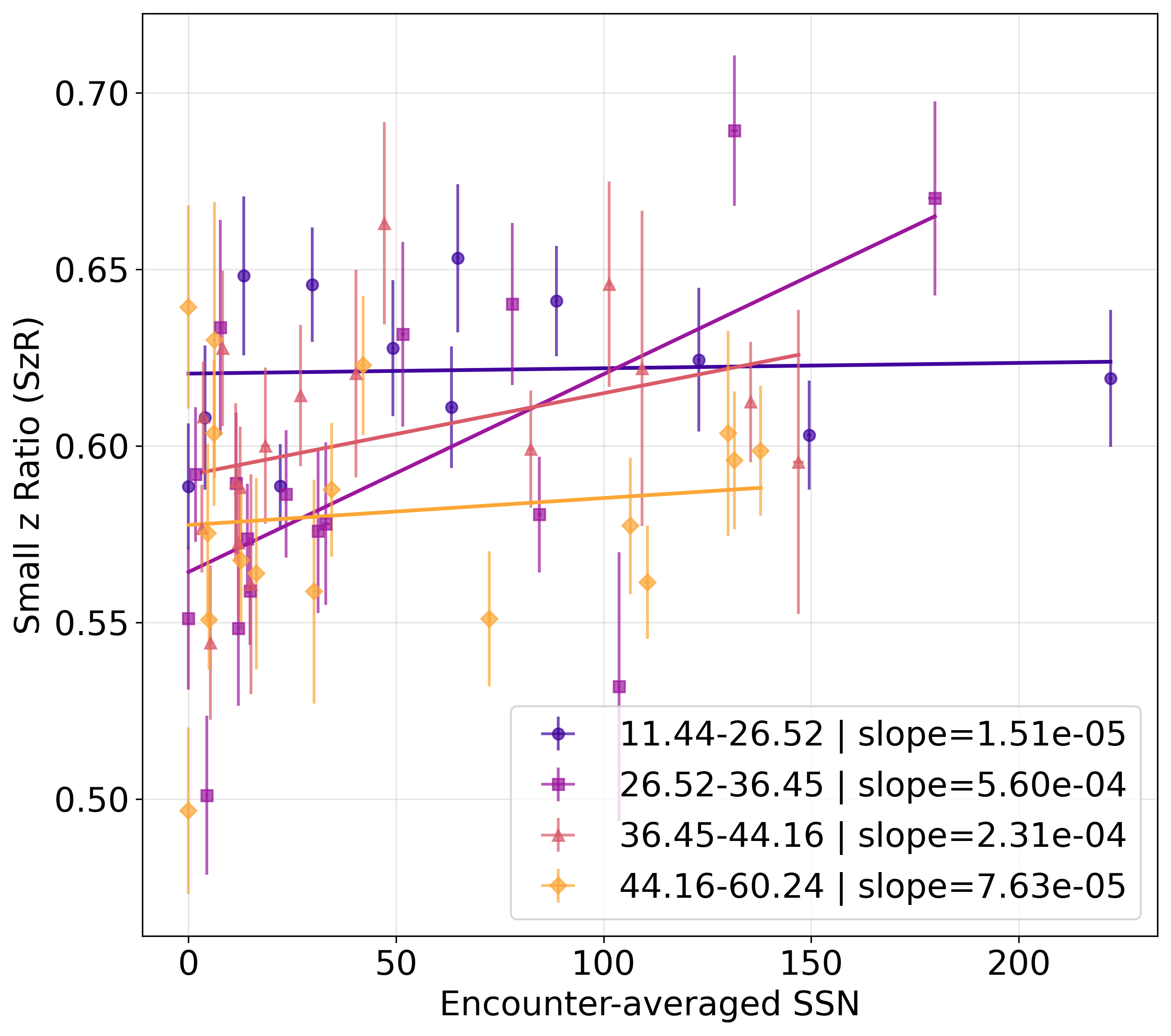}
\caption{Evolution of the Small z Ratio ($SzR$) with sunspot number (SSN) across the solar cycle starting from low solar activity to increase towards the peak, binned in distances in colours shown in the legend. Linear fitting is also performed, the slopes are noted in the legend.}
\label{fig:$SzR$_time_SSN}
\end{figure}

To help us disentangle the dependence of the deflection on the level of solar activity and on the distance, we show in Fig.~\ref{fig:AS1_phase_distance} the $SzR$ ratio versus distance and SSN, both of which are binned. We use equispaced bins for the distance, and three bins for the SSN (0-12, 12-55, 55-250), based on the histogram shown in the bottom right panel of Fig.~\ref{fig:description_of_dataset}. Each bin {of the marginal distribution} typically contains between 2000 and 4000 values to guarantee the statistical robustness of the results, further shown in Fig.~\ref{fig:sanity_checks}. This means that small changes in $SzR$ are meaningful, even though $SzR$ ranges from 0.49 to 0.68 only. 

Figure~\ref{fig:AS1_phase_distance} confirms again the gradual decrease of the $SzR$  with larger distances from the Sun. This result suggests that the fraction of large deflections increases at larger distances. 

Interestingly, Fig.~\ref{fig:AS1_phase_distance} also suggests that there is a solar cycle dependence in the angular deflection of switchbacks, since small {values of $SzR$} are relatively more frequent near solar minimum. To our knowledge, this is the first observation of such a solar cycle dependence in the properties of switchbacks.

For a more quantitative understanding of the behaviour of $SzR$ through the solar cycle, we plot in Fig.~\ref{fig:$SzR$_time_SSN} the encounter-averaged $SzR$ with its standard error versus distance from the Sun. Values are binned in distance from the Sun. 
The error bars are defined as standard error, which is standard deviation in the given bin normalised to the square root of the number of data points in the given bin. This figure confirms the tendency of $SzR$ to increase with the level of solar activity, i.e. large deflections are less frequent when the Sun is active.

\vspace{-15pt}
\subsection{Multilinear regression on $z$}
\label{Sec:multilinear}

The main challenge we face is determining the dependence of the deflection angle on the sunspot number, given that this angle is also influenced by quantities (solar wind speed, distance from the Sun, Alfvén mach number, etc.) that all depend on each other. We are dealing here with an attribution problem, in which one wants to determine the dependence of a given quantity ($z$ or $SzR$) on one single regressor only (SSN), while excluding the contribution of the others. To put the results of Fig.~\ref{fig:AS1_phase_distance} on firmer ground and separate the individual contributions, we perform ordinary least squares (OLS) regression on the defined quantity $SzR$, using as regressors SSN and the distance from the Sun ($R_{\odot}$). 

In the following, we use a 20-hour rolling window to calculate the $SzR$ (as explained further in Appendix~\ref{AppendixA}) to find the following expression for the linear model that relates the $SzR$ to the two variables

\vspace{-7pt}
\begin{equation} \label{equation:multilinear_regression}
SzR = 0.049 + 1.2\times10^{-5}(\mathrm{SSN}) + 0.00038(R_{\odot}).
\end{equation}

The standard errors of the model coefficients are presented in Appendix~\ref{AppendixA}). The positive SSN coefficient is consistent with the trends shown in Fig.~\ref{fig:$SzR$_time_SSN}.

Note that Pearson's correlation coefficient between the observed and modelled value of $SzR$ is low ($R^2 = 0.023$) because of the large scatter in the values of $SzR$. For that reason, our model should really be considered as a means for detecting a functional relationship, and not as an attempt to predict the value of $SzR$.

The key result is the positive value of the model coefficient for the sunspot number, which is statistically significant at a 99.9\% level. The model fit confirms the trend seen in Fig.~\ref{fig:AS1_phase_distance}, showing that the probability of large switchbacks decreases with a more active Sun, and increases farther away from the Sun.

This result is further supported by multilinear fits with different rolling window sizes for $SzR$ calculations (Table~\ref{tab:Multilinear_regression_coefficients}), where the dependence on solar activity and distance remain consistently positive, supporting our claims further.

\vspace{-15pt}
\section{Discussion and Conclusion}
\label{Sec:Discussion}

Switchback deflection properties, quantified through $SzR$, exhibit a statistically significant but weak dependence on solar activity. For showing this, we introduced a quantitative framework combining radial binning, sunspot-based activity classification, and regression analysis, enabling a first order separation of heliocentric distance and solar activity effects. This approach isolates weak trends that would otherwise be obscured by the strong coupling between distance and cycle phase in PSP observations. 
 
The non-uniform sampling across PSP encounters introduces an uneven coverage of our observables, which may reduce the statistical robustness of the inferred trends. This can be tested in different ways; in Appendix~\ref{AppendixA} we provide one piece of evidence to support our conclusions.

The dependence on the solar cycle is likely to be the consequence of several mechanisms, some of which may cancel each other out, hence the weak signature. The most immediate contribution comes from a modulation of the coronal sources. Solar cycle variations in coronal magnetic topology, including changes in the distribution and geometry of coronal holes and active regions, are known to modulate the availability of open magnetic flux and the solar wind properties \citep{Schwenn2006,Abbo2016}. In particular, the fractional coverage of coronal holes and active regions, as well as the complexity of open–closed flux boundaries, may influence the occurrence of interchange reconnection \citep{Fisk2005} and the subsequent injection of Alfv\'enic perturbations into the solar wind \citep[e.g.][]{Drake2021, 2024A&A...692A..71T}. Global changes in coronal topology over the solar cycle are therefore expected to modulate the generation of switchbacks and potentially affect their occurrence rate. However, it remains to be investigated how it could change the amplitude of the injected Alfvénic perturbations. Finally, these connections remain indirect, and a direct observational link between solar-cycle evolution of coronal structure and switchback properties has not yet been firmly established \citep{2026SSRv..222...14B,2026SSRv..222...43W}.

A different contribution comes from changes in the in situ processes, which are expected to substantially modify switchback properties as the solar wind expands \citep{2025A&A...700A..51B,payne_evolution_2026}. Nonlinear evolution of Alfv\'enic fluctuations, turbulent cascade, and geometric expansion can distort magnetic deflections with distance \citep{Matteini2007,Bruno2013,Mallet2021}. Recent modelling efforts suggest that both in situ generation and coronal release of structured magnetic fields may contribute to the observed switchback population \citep[e.g.][]{2024A&A...692A..71T}. In addition, solar-cycle variations in the location and morphology of the Alfv\'en surface, whose height, thickness, and asphericity evolve with activity \citep{Badman_2025}, could alter the conditions under which switchbacks may form and/or propagate \citep{Shi2024, Touresse2026}. This introduces an additional layer of activity-dependent control, although the increasing geometric and dynamical complexity of the Alfv\'en surface toward solar maximum may also act to smear out clear cycle-dependent signatures. Finally, we note that simulation studies by \citet{Touresse2026} have shown that the deflection may depend on the way the 3D structure is crossed by the spacecraft.

It is very likely that these different contributions, taken individually, should lead to a stronger solar cycle dependence of the deflection angles. The weak response we observe may therefore be the result of competition between the various mechanisms. This supports the hypothesis that switchbacks cannot be explained by purely coronal or in situ mechanisms alone but instead reflect a coupled process in which several mechanisms are important. This seems to be especially true for switchbacks that have a large deflection.

More broadly, these results show that solar activity influences not only bulk solar wind properties but also embedded magnetic structures and Alfv\'enic fluctuations \citep{Bruno2013,Owens2013}. The low correlation $R^2$ between the modelled and observed $SzR$ indicates that the sunspot number and the heliocentric distance account for only a small fraction of the observed variability, highlighting the need for multi-parameter analyses. Future work should incorporate additional plasma parameters such as solar wind speed, Alfv\'enicity, and plasma $\beta$. Such approaches are essential to help disentangle coronal source effects from in-situ evolution and to establish a comprehensive picture of switchback formation across the solar cycle.

\begin{acknowledgements}
This work was supported by the CEFIPRA Research Project No. 6904-2. T.D, C.F. and G.S. acknowledge the financial support of the Centre National d’Études Spatiales (CNES), France (ROR: \url{https://ror.org/04h1h0y33}), within the framework of the Parker Solar Probe mission, and of the Action Thématique Soleil-Terre (ATST) of CNRS/INSU PN Astro, co-funded by CNES and CEA. VU acknowledges NASA for support under award number 80NSSC25K7956. Sunspot data from the World Data Center SILSO, Royal Observatory of Belgium, Brussels, \citep{SILSO_Sunspot_Number}. This research utilised the Python libraries matplotlib \citep{2007CSE.....9...90H}, seaborn \citep{Waskom2021} and the NumPy computational environment \citep{citeulike:9919912}. 

\end{acknowledgements}

\vspace{-15pt}
\bibliographystyle{aa}
\bibliography{refer}

\begin{appendix}
\section{On the Robustness of the Results}
\label{AppendixA}

\begin{figure}[h!]
\centering
\includegraphics[width=0.5\textwidth, keepaspectratio]{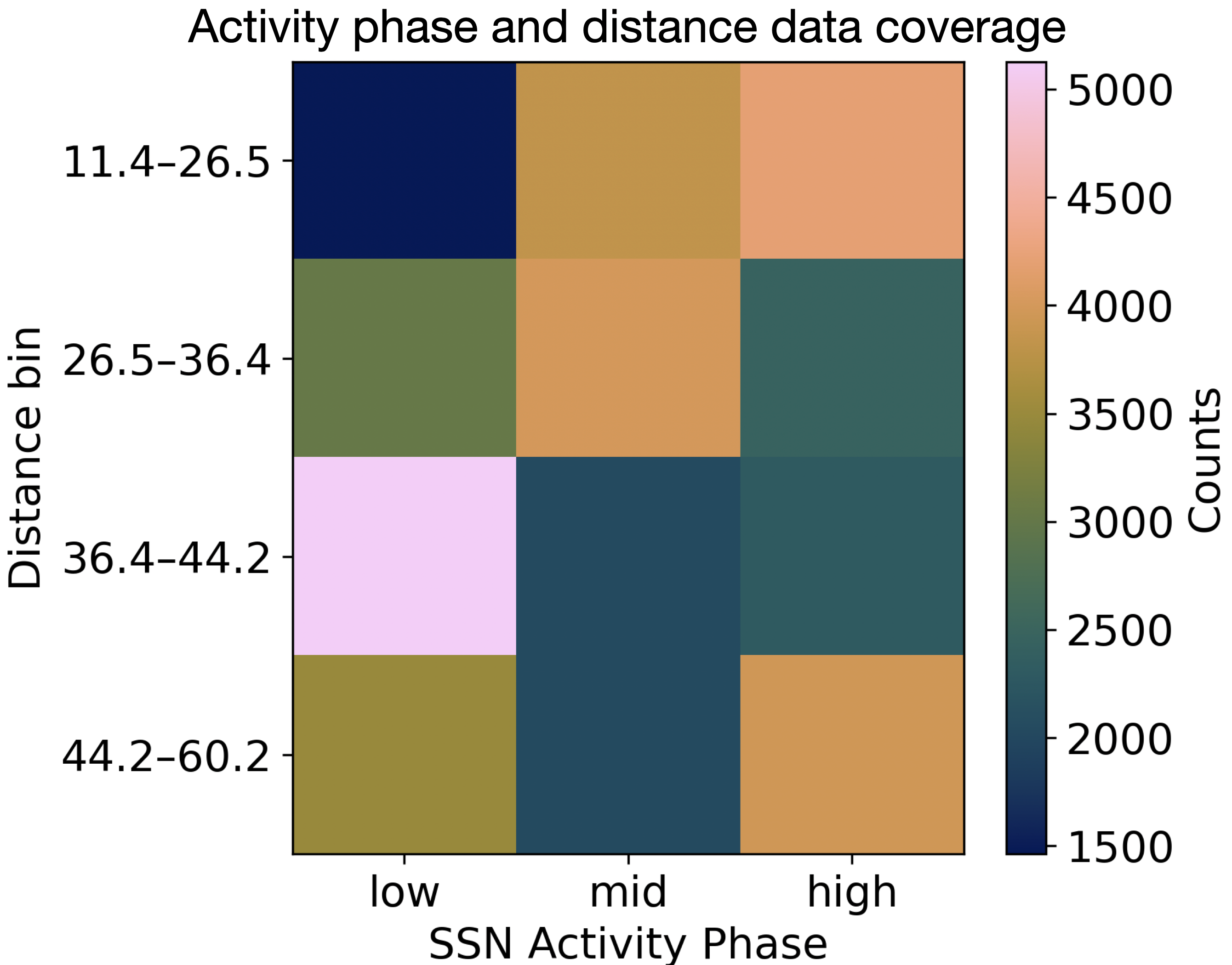}
\caption{Solar activity phase and distance data coverage showing that there no phase and no distance}
\label{fig:sanity_checks}
\end{figure}

We performed a series of robustness tests to verify the stability of the inferred radial and solar-cycle dependencies. Alternative radial binning schemes (fixed-width bins instead of equiprobable) yields consistent behaviour as shown in Fig.~\ref{fig:sanity_checks}, confirming that the observed trends are not artefacts of binning or uneven sampling.

We performed an ordinary least squares (OLS) regression on the derived parameter $SzR$ (Sect.~\ref{Sec:multilinear}) using the sunspot number (SSN) and the heliocentric distance ($R_\odot$) as independent variables.  At the 5\% significance level, all coefficients are significantly different from zero, supporting the statistical validity of the model.

\begin{figure}[h!]
\centering
\includegraphics[width=1.0\textwidth, height = 0.4\textheight, keepaspectratio]{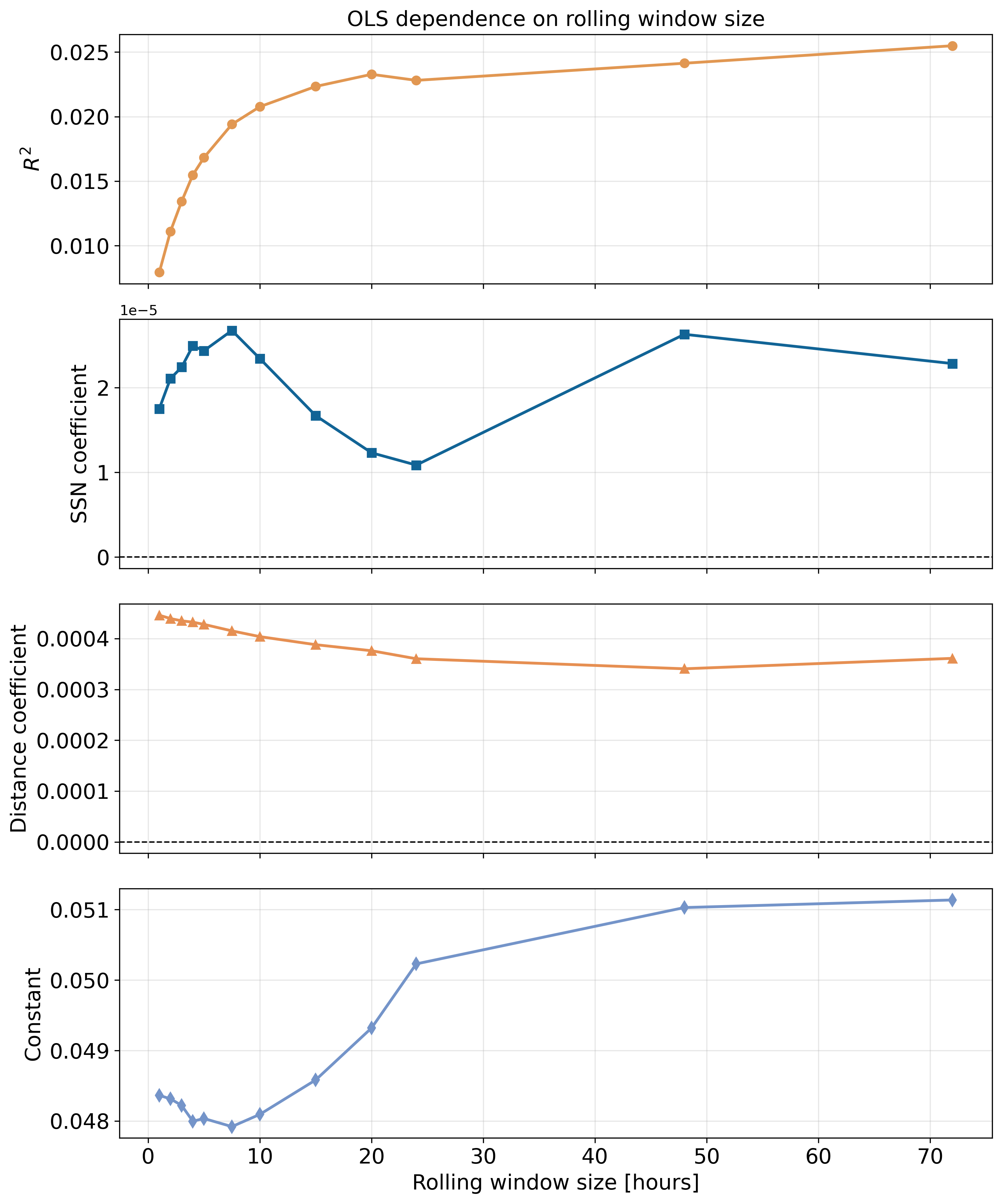}
\caption{The dependence of rolling window size on the parameters of OLS regression: $R^2$, SSN coefficient, Distance coefficient and the constant.}
\label{fig:OLS_Rolling_Window}
\end{figure}

We computed the parameter $SzR$ using a 20-hour rolling window. A rolling window approach was preferred over discrete binning because switchback timestamps correspond to event start times rather than a continuous time series, which would lead to uneven temporal bins. Furthermore, discrete binning would require averaging the associated distance and SSN values within each bin, potentially reducing the information content of the data. Figure~\ref{fig:OLS_Rolling_Window} and Table~\ref{tab:Multilinear_regression_coefficients} show the dependence of the OLS coefficients and the coefficient of determination ($R^2$) on the selected window size. The regression coefficients reach a stable plateau for window sizes of approximately 20 hours and larger. We therefore adopt a 20-hour window throughout the analysis presented in Sect.~\ref{Sec:multilinear}.

\begin{table}[h]
\begin{center}
\begin{tabular}{|p{0.5cm}|p{1.1cm}|p{1.1cm}|p{0.8cm}|p{1.0cm}|p{1.5cm}|p{1.1cm}|}
\hline
win & $R^2$       & Intercept    & SSN  & Distance  & P value         & P value  \\
dow &       &     &   $\times 10^{-5}$ &   $\times 10^{-3}$ &  SSN      & Distance  \\
\hline
1H     & 0.00792 & 0.0483 & 1.8  & 0.446       & 1.61e-22  & 0.0         \\
2H     & 0.01110 & 0.0483 & 2.1  & 0.439       & 2.68e-45  & 0.0         \\
3H     & 0.01342 & 0.0482 & 2.2  & 0.435       & 1.29e-62  & 0.0         \\
4H     & 0.01547 & 0.0479 & 2.5  & 0.432       & 2.26e-89  & 0.0         \\
5H     & 0.01682 & 0.0480 & 2.4  & 0.428       & 6.01e-95  & 0.0         \\
7.5H   & 0.01940 & 0.0479 & 2.7  & 0.415       & 3.79e-139 & 0.0         \\
10H    & 0.02077 & 0.0480 & 2.3  & 0.404       & 1.56e-122 & 0.0         \\
15H    & 0.02234 & 0.0486 & 1.7  & 0.388       & 1.33e-74  & 0.0         \\
\textbf{20H}    & \textbf{0.02328} & \textbf{0.0493} & \textbf{1.2}  & \textbf{0.376}       & \textbf{3.26e-46}  & \textbf{0.0}        \\
24H	&0.02281	&0.0502	&1.1	&0.360	&6.13e-39	&0.0 \\
48H	&0.02413	&0.0510	&2.6	&0.341	&1.29e-283&	0.0\\
72H	&0.02549	&0.0511	&2.3	&0.361	&1.72e-230&	0.0\\
\hline
\end{tabular}\caption{The multilinear regression coefficients and P values of the Distance and SSN slopes for different rolling windows in the analysis presented in Sec.~\ref{Sec:multilinear}. Window of 20H is shown in boldface for reference.}
\label{tab:Multilinear_regression_coefficients} 
\end{center}
\end{table}

The choice of rolling-window size does not affect our conclusions regarding the solar cycle dependence of $SzR$. Although Pearson's correlation coefficient $R^2$ increases for larger window sizes owing to the reduced effective sample size, the signs and overall magnitudes of the SSN and distance coefficients remain stable. This demonstrates that the inferred dependence of $SzR$ on solar activity and heliocentric distance is robust against the choice of window size. The consistently positive value of the coefficient for SSN indicates that increasing solar activity is associated with a smaller fraction of large-angle deflections in the in situ observations.

\end{appendix}

\end{document}